


\input phyzzx
\hoffset=0.2truein
\voffset=0.1truein
\hsize=6truein
\def\TITLEPAGE{\frontpagetrue}
\def\IC#1{\hbox to\hsize{\tenpoint \baselineskip=12pt
	\hfil\vtop{\hbox{\strut IC-94-#1}
	\hbox{\strut Research and}
	\hbox{\strut Development Report}}}}
\def\cinve#1{\hbox to\hsize{\tenpoint \baselineskip=12pt
	\hfil\vtop{\hbox{\strut CINVESTAV-FIS-#1}
	\hbox{\strut Report}
	\hbox{\strut }}}}

\def\CIEA{\smallskip
	\address{Departamento de Fisica,\break 
          Centro de Investigaci\'on y Estudios Avanzados del IPN,\break
Apdo. Postal 14-740, c.p. 07000 M\'exico, D.F., MEXICO}}

\def\AUTHOR#1{\vskip .5in \centerline{#1}}

\def\ABSTRACT#1{\vskip .5in \vfil \centerline{\twelvepoint \bf Abstract}
	#1 \vfil}
\def\ENDTITLEPAGE{\vfil\eject\pageno=1}

\def\sqr#1#2{{\vcenter{\hrule height.#2pt
      \hbox{\vrule width.#2pt height#1pt \kern#1pt
        \vrule width.#2pt}
      \hrule height.#2pt}}}

\def\section#1#2{
\noindent\hbox{\hbox{\bf #1}\hskip 10pt\vtop{\hsize=5in
\baselineskip=12pt \noindent \bf #2 \hfil}\hfil}
\medskip}

\def\underwig#1{	
	\setbox0=\hbox{\rm \strut}
	\hbox to 0pt{$#1$\hss} \lower \ht0 \hbox{\rm \char'176}}

\def\bunderwig#1{	
	\setbox0=\hbox{\rm \strut}
	\hbox to 1.5pt{$#1$\hss} \lower 12.8pt
	 \hbox{\seventeenrm \char'176}\hbox to 2pt{\hfil}}


\def\PRL#1#2#3{{\it Phys. Rev. Lett.} {\bf#1} (#2) #3}
\def\NPB#1#2#3{{\it Nucl. Phys.} {\bf B#1} (#2) #3}

\def\PRD#1#2#3{{\it Phys. Rev.} {\bf D#1} (#2) #3}

\def\PLB#1#2#3{{\it Phys. Lett.} {\bf #1B} (#2) #3}
\def\JMP#1#2#3{{\it J. Math. Phys.} {\bf #1} (#2) #3}

\def\AoP#1#2#3{{\it Ann. Phys.} {\bf #1} (#2) #3}

\def\IJMPA#1#2#3{{\it Int. J. Mod. Phys.} {\bf A#1} (#2) #3}

\def\MPLA#1#2#3{{\it Mod. Phys. Lett.} {\bf A#1} (#2) #3}        

\def \hep{hep-th/}


\def \od {O(d-1, 1)}
\def \odp {O(d + p -1, 1)}
\def \dm{\partial_{\mu}}


\REF\gibma{G.W. Gibbons and K. Maeda, \NPB{298}{1988}{741}.}

\REF\GaHS{D. Garfinkle, G. Horowitz and A. Strominger,
\PRD{43}{1991}{3140}, erratum \PRD{45}{1992}{3888}.}

\REF\HS{G. Horowitz and A. Strominger, \NPB{360}{1991}{197}.}

\REF\HoHSt{J. Horn, G. Horowitz and A. Steif, \PRL{68}{1992}{568}.}

\REF\HoH{J. Horn and G. Horowitz, \PRD{46}{1992}{1340}.} 

\REF\CKO{B. A. Campbell, N. Kaloper and K. A. Olive,  
\PLB{285}{1992}{191}.}

\REF\BUSCHER{T. Buscher, \PLB{194}{1987}{59}; \PLB{201}{1988}
{466}.}

\REF\STW{A. Shapere, S. Trivedi, and F. Wilczek, \MPLA{6}{1991}
{2677}.}

\REF\NSW{K. S. Narain, M. H. Sarmadi, and E. Witten, \NPB
{279}{1987}{369}.}

\REF\CARA{J. Cardy, \NPB{205}{1982}{17}.}

\REF\SW{A. Shapere and F. Wilczek, \NPB{320}{1989}{669}.}

\REF\Duali{A. Giveon, E.
Rabinovici, and G. Veneziano, \NPB{322}{1989}{167}; 
A. Giveon, N. Malkin,
and E. Rabinovici, \PLB{220}{1989}{551}; 
W. Lerche, D. L\"ust, and N.
P. Warner, \PLB{231}{1989}{417}.}

\REF\MV{ K. A. Meissner and G.
Veneziano, \PLB{267}{1991}{33} and \MPLA{6}{1991}{3397}; 
M. Gasperini, J. Maharana, and G. Veneziano \PLB{272}{1991}{277}.} 

\REF\JMJS{J. Maharana and J. H. Schwarz, \NPB{390}{1993}{3},
(\hep 9207016).}

\REF\hasen{S. F. Hassan and A. Sen, \NPB{375}{1992}{103}. 
(\hep 9109038).}
 
\REF\sentre{A. Sen, \PRL{69}{1992}{1006}. (\hep 9204046).}

\REF\sencua{A. Sen, \NPB{404}{1993}{109}. (\hep 9207053).}

\REF\Smarcus{J. H. Schwarz and N. Marcus, \NPB{228}{1983}{145}.}

\REF\sencin{A. Sen, \IJMPA{9}{1994}{3707}. (\hep 9402002).}

\REF\sensei{A. Sen, \NPB{440}{1995}{421-440}. (\hep 9411187).}

\REF\GK{D. Galtsov and O. Kechkin, \PRD{50}{1994}{7394}, (\hep 9407155).}

\REF\asen{A. Sen, \NPB{447}{1995}{62-84},  (\hep 9503057).}

\REF\KURA{A. Kumar and K. Ray, \NPB{453}{1995}{181-198}, (\hep 9506037).
\PLB{358}{1995}{223-228}, (\hep 9503154).}

\REF\ELHA{J. Ehlers, in Les theories Relativistes
 de la Gravitation, CNRS, Paris, 1959, pag. 275; B. K. Harrison, 
\JMP{9}{1968}{1774}.}

\REF\Gero{R. Geroch, \JMP{13}{1972}{394}.}

\REF\KCH{W. Kinnersley and D. M. Chitre, \JMP{19}{1978}{2037-2042}.}

\REF\KKOT{R. Kallosh, D. Kastor,  T. Ort{\'\i}n and T.  Torma, 
\PRD{50}{1994}{6374-6384}, (\hep 9406059).}

\REF\cliffo{C. V. Johnson and R.C. Myers,\PRD{50}{1994}{6512}, (\hep 
9406069.)} 

\REF\montonen{C. Montonen and D. Olive, \PLB{72} {1977}{117}.}

\REF\PLE{J. F. Pleba\'nski, \AoP{90}{1975}{196}.}

\REF\PD{J. F. Pleba\'nski and M. Demianski, \AoP{98}{1976}{98}.}

\REF\GGK{A. Garc{\'\i}a, D. Galtsov and O. Kechkin, 
\PRL{74}{1995}{1276}.}

\REF\KW{W. Kinnersley and M. Walker, \PRD{2}{1970}{1359}.}

\REF\DGKT{F. Dowker, J. P. Gauntlett, D. A. Kastor and J. Traschen, 
\PRD{49}{1994}{2909-2917}, (\hep 9309075).}


\TITLEPAGE

\cinve{10/96}

\bigskip           

\titlestyle {Boosting some type--D metrics 
\foot{Work partially supported by Conacyt-Mexico}}

\AUTHOR{L. Palacios}

\CIEA

\ABSTRACT{We are presenting a general solution to the classical 
Einstein--Maxwell--dilaton--axion equations starting from a 
metric of type--D. Namely, this
stringy solution is the result of a transformation on a general 
vacuum type--D solution to the Einstein's equations 
which was studied in detail some years ago.}

\medskip

\ENDTITLEPAGE

\eject


\noindent {\bf 1.  Introduction}

There has been a good deal of work on several topics around 
the low energy
effective action of string theory. In particular, the analysis 
of the equations
of motion of the dilaton and/or axion -- graviton  system
has given rise to solutions that exhibit some new properties
of the black holes (or black p-branes) of this string gravity [\gibma - 
\CKO].
One of the first important analyses of the classical dilaton--graviton 
system was reported in [\gibma], which shows that the physical 
properties of the black 
hole solutions vary depending on the value of the coupling 
constant of the dilaton field
(e.g. black hole thermodynamics, stability of inner horizons and so on).

Similar work for the dilaton -- gravity system 
has been presented in [\GaHS] and extended in [\HS]   
to higher dimensions with solutions associated with the so-called black 
strings and black p-branes.
An interesting approach was adopted in [\HoHSt] while studying 
the axion-gravity system, using an embedding--boosting--duality  
sequel of steps enabling generation of an isometry, a conserved 
momentum in the translation invariant direction and finally by duality 
[\BUSCHER] an axion charge that can be associated with the 
conserved momentum.

Later, an important step in the construction of dyonic solutions of the
string equations was given in  
[\STW], in which the exact SL(2, R) symmetry of the equations of motion 
of the low energy effective string action in four dimensions was used
to construct a dual solution with electric and magnetic charge.

A systematic method for studying the space of solutions of the equations
of motion of the LEEA of the heterotic string is given in [\hasen] (see 
also refs. therein).
 The main ingredients that Sen uses are a realisation of the 
$O(d, d+p)$ duality symmetry [\NSW, \SW, \Duali,  \MV]
and elements of the subgroup $\od \times \odp$ $\subset$ $O(d, d+p)$ 
to generate the new solutions. This kind of symmetry arises in the 
process of compactification. For example,
the 10-dimensional $N=1$ supergravity with $p$ abelian 
vector supermultiplets (in addition to the supergravity multiplet) is 
dimensionally reduced to
$10 - d $ dimensions with a global $O(d, d+p)$ 
symmetry [\JMJS, see refs therein].

The $O(7, 23)$ group has been used to generate charged solutions [\sentre\ - 
\sensei] from a neutral solution with  time independent field 
configurations including  the moduli fields.
The $O(8, 24)$ group of symmetries of string theories was studied and used 
 to generate a general charged solution [\Smarcus, \sensei]. 
The case with $SO(2, 3)$ as the symmetry group 
has been studied in [\GK] in which the moduli fields are not considered 
and an equivalent action that reproduces the equations of motion of the LEEA has 
been 
used. There is a further reduction to two-dimensions that leads to an infinite 
dimensional symmetry $\widehat {O(8, 24)}$ [\asen, \KURA, see refs. therein]. All these 
symmetries 
can be seen as the extensions of the symmetries studied in the Einstein-Maxwell
theory such as Ehlers-Harrison transformations, the Geroch group and 
the Kinnersley-Chitre symmetries [\ELHA, \Gero, \KCH]. 

A realisation of the $O(6, 22)$ symmetry is given by the LEEA in 
four dimensions 
$$S = \int d^4 x \sqrt{-g}(R(g) - \half (\nabla\Phi)^2 - 
		  {1\over 12}  e^{-2 \Phi}  H^2 - 
		  {1\over 8}   e^{- \Phi} F^2 + ...)\eqn\action$$
where $\Phi$ is the dilaton field, $F_{\mu\nu}$ is the Maxwell field 
associated with a $U(1) \subset E_8 
\times E_8 $ gauge field, $g_{\mu \nu}$ is the Einstein metric that is 
related to 
the string metric as follows $g_{\mu\nu} = e^{-\Phi} G_{\mu\nu}$.
The three-form with only the Chern-Simons term for the $U(1)$ gauge field 
is given by 
$$H_{\mu\nu\rho} = \dm B_{\nu\rho} - {1\over 4} A_\mu F_{\nu\rho} +
cyclic\;permutations\eqn\lache$$
where $B$ is the antisymmetric tensor gauge field.

Considering only one copy of the $U(1)$ gauge field,
the background field configuration  
can be contained in the $9 \times 9$ matrix  $M$ as follows [\sentre]
$$M = \left(\matrix{K_-^T G^{-1}K_- &K_-^T G^{-1}K_+ &-K_-^T G^{-1}A \cr
		    K_+^T G^{-1}K_- &K_+^T G^{-1}K_+ &-K_+^T G^{-1}A \cr
  -A^T G^{-1}K_- &-A^T G^{-1}K_+  &A^T G^{-1}A 
\cr}\right)\eqn\laomega$$
where 
$$K_{\pm} = - B - G - {1\over 4}AA^T \pm \eta_4$$
with $\eta_4 = diag(+, +, +, -)$.

A new field configuration can be obtained from the following transformation [\sentre] 
$$M'(G', B', A') = \Omega M \Omega^T \eqn\newconfi$$
 and the dilaton field transforms under the duality rules [\BUSCHER]
$$\Phi ' = \Phi - \half \log ({Det G\over Det G'})\eqn\dilato$$
with $\Omega$ given by
$$ \Omega = \left(\matrix{I_7&0 &0\cr
			   0&\cosh \alpha & \sinh\alpha \cr
			   0&\sinh\alpha & \cosh \alpha \cr}\right)\eqn\laomega$$
where $I_7$ is the identity matrix $7 \times 7$.
Then, one can generate (in general) inequivalent backgrounds which are time 
independent [\sencua] as this subgroup leaves invariant the equations 
of motion.
We can start from a given solution of the pure Einstein's equation (i.e. $A_\mu = 
B_{\mu\nu} = \Phi =0 $) and construct a solution of the full string theory 
equations of motion.

In addition to the $O(d, d+p)$ symmetries, 
there is also an exact $SL(2,R)$ symmetry of the equations of motion of the 
LEEA of the heterotic string in four dimensions  that enables construction of 
dyonic solutions starting from one configuration with either 
an electric or magnetic charge. 
The equations of motion of the action \action\ are invariant under a
$SL(2, R)$ duality symmetry of the type [\STW, \sencua]
$$ F_- \rightarrow - \bar\lambda F_-,  \qquad F_+ \rightarrow - \lambda 
F_+\eqn\slf$$
$$ \lambda \rightarrow \lambda + c, \qquad \lambda \rightarrow - 
1/\lambda\eqn\sllam$$
with
$$F^{\mu \nu}_{\pm} = F^{\mu \nu} \pm i \tilde F^{\mu \nu}$$
$$ \lambda = \Psi + i e^{-\Phi}\eqn\coupling$$
the imaginary part of $F$ is
$$ \tilde F^{\mu \nu} = \half {\epsilon^{\mu\nu\rho\sigma}\over \sqrt{-g}}
F_{\rho\sigma}$$
The real part of the modular parameter $\lambda$ is
given by
the axion field and its imaginary part is given by the dilaton field.
This can be seen as one of the manifestations of 
the S-duality symmetry in the LEEA [\montonen, \sencin].
The SL(2, Z) symmetry has also appeared in lattice gauge theories 
with a theta-term [\CARA, \SW], where
the models without $\theta$-term are self-dual with respect to 
inversion of the coupling constant and with a $\theta$-term this duality may be extended
to an invariance $D : \zeta \rightarrow - 1/ \zeta,  T: \zeta \rightarrow  \zeta  + 1 $
where $\zeta $ is a complex coupling constant. This kind of duality resembles the 
transformation \sllam\ of the  
coupling \coupling\ in the case of the heterotic LEEA [\STW].
 
In section 2, we shall briefly review the most general type D metric and 
 construct a solution to the 
heterotic string equations starting from a generic type D metric which is a solution to 
the Einstein equations. The method uses an $O(1, 1)$ transformation and 
only one copy of a $U(1)$ gauge field. Next,  
we shall use the SL(2,R) transform to generate the dual solution. We should 
like to 
investigate the symmetric role played by the parameters and the coordinates on which the 
field configuration depends. Finally, we shall present the string version of the 
C-metric [\KW].

One should note that 
the $SL(2, R)$ transformation together with the $O(7, 23)$ generate
the $O(8, 24)$ group.
Then, using an element of $O(8, 24)$ with one single transformation, it 
is possible to generate a 59 parameter solution 
provided that one is transforming a time independent configuration 
[\sencin, \sensei], then 
giving rise to solutions that contain the previous solutions
discussed in [\GaHS, \HoH,  \sentre, \KKOT, \cliffo].

\vfil\eject

\noindent{\bf 2. Type--D solutions yesterday and today}

In solving the Einstein--Maxwell equations 
the most general type--D metric has been given in [\PD] and reads
as follows 
$$ \eqalign{ds^2 = {1\over (1 - pq)^2} &[{p^2 + q^2\over P} dp^2 + 
{P\over p^2 + q^2}(d\tau + q^2 d\sigma)^2 \cr
&+ {p^2 + q^2\over Q}dq^2 - {Q\over p^2 + q^2}(d\tau - p^2 
d\sigma)^2]\cr}\eqn\pdmetri$$
These space-times are described by some
real coordinates $x^\mu = ( p, q, \sigma, \tau ) $
and with  signature $(+, +, +, -)$. $P(p)$
and $Q(q)$ are quartic polynomials 
$$P(p) = (-(\lambda /6) - g_0^2 + \gamma) + 2np - \epsilon p^2 +
 2m p^3 + (-(\lambda /6) - e_0^2 - \gamma) p^4\eqn\ppoly$$
and $$Q(q) = - P(-q) \eqn\qpoly$$
The parameters $m, n, e_0, g_0$ (mass, NUT parameter, electric and
magnetic charge) are the 
dynamical parameters 
that enter into the electromagnetic field and the curvature. The 
parameters $\gamma$ and $\epsilon$ are kinematical parameters related to
the rotation and to some constant acceleration of a particle with mass 
$m$. The parameter $\lambda$ is the cosmological constant.

The electromagnetic field is given by 
$$ f_+ \equiv (f_{\mu\nu} + i \tilde f_{\mu\nu}) dx^\mu dx^\nu 
= - d ({e_0 + i g_0\over q + i p } (d\tau - i pq d\sigma ))\eqn\einelec$$
with $\tilde f^{\mu\nu} = \half (-g)^{-1/2} \epsilon^{\mu\nu\rho\sigma}
f_{\rho\sigma}$.  The complex invariant
${\cal F} = \half (f_{+\mu\nu})^2 = f_{\mu\nu}f^{\mu\nu} + i 
f_{\mu\nu}\tilde f^{\mu\nu}$ is  
$${\cal F} = -\half(e_0 + i g_0)^2({1 - pq\over{q + i p}})^4\eqn\invaele$$
and the only nonvanishing component of the Weyl 
tensor is 
$$C^{(3)} = - 2 ({1 - pq\over q + i p })^3 [(m + in)  -
   ( e_0^2 + g_0^2) {1 + pq\over q - i p}]\eqn\weyl$$ 
It would be interesting to see 
the extension of this general metric in the context of string theory. For this, we will first use a 
generic type-D neutral solution to generate a charged solution to the equations of motion of the 
action \action, that includes the dilaton and axion fields.

\smallskip

\noindent {\bf 2.1 Type D metrics without conformal factor and their stringy 
version}

From \pdmetri\ one obtains a class of solutions where the conformal  factor 
has been removed [\PLE]. That is equivalent to eliminate 
the parameter of acceleration
$$ \eqalign{ds^2 &= {p^2 + q^2\over P} 
dp^2 + {P\over p^2 + q^2}(d\tau + q^2 d\sigma)^2 \cr
&+ {p^2 + q^2\over Q}dq^2 - {Q\over p^2 + q^2}(d\tau - p^2 
d\sigma)^2\cr}\eqn\sinconformal$$
with the structural functions given by
$$P(p) = \gamma - g_0^2 + 2np - \epsilon p^2 - (\lambda /3) 
p^4\eqn\pscf$$
$$Q(q) = \gamma + e_0^2 - 2mq + \epsilon q^2 - (\lambda /3)
q^4\eqn\qscf$$    
This solution contains six continous and a discrete parameters.
From which $\epsilon$ can take the values  $\pm 1, 0$. 
The complex invariant is given by
$${\cal F} = -\half(e_0 + i g_0)^2{1\over(q + i p)^4}\eqn\ncfinv$$
The only nonvanishing component of the Weyl tensor
$$C^{(3)} = {-2\over (q + i p)^3} [(m + in)  -
   ( e_0^2 + g_0^2) {1 \over q - i p}]\eqn\ncfweyl$$
We consider the neutral solution \sinconformal\ - \qscf\ with  $ e_0 = 
g_0 = 0$ to generate a charged 
solution by using an $O(1, 1)$ transformation given by $\Omega$ in \newconfi\ -  
\laomega. From the properties of the starting metric we should obtain an stringy metric with all the 
known parameters except the acceleration one. The new charges (electric and magnetic) will be 
consistent with the presence of the dilaton and axion fields.

The new metric $ds'^2$, the electromagnetic gauge field $A_\mu$
and the 
antisymmetric field $B_{\mu\nu}$ are obtained according to \newconfi. The 
dilaton field is obtained from \dilato,
thus, the resulting new Einstein metric given by 
$ds_E^2 = e^{-\Phi} ds'^2$ is
$$ \eqalign{ds_E^2 &= {D\over P}
dp^2 + {P\over D}(d\tau + (b_1 q^2 + b_0 Q) d\sigma)^2 \cr
&+ {D\over Q}dq^2 - {Q\over D}(d\tau - (b_1 p^2 - b_0 P)
d\sigma)^2\cr}\eqn\chargesconf$$
with 
$$D = b_1 (p^2 + q^2) - b_0 (P - Q)$$and
$$ b_0 = {(1 - \cosh \alpha)\over 2};
\qquad b_1 = {(1 +\cosh \alpha)\over 2}$$
The scalar dilaton field, according to \dilato\ is given by
$$e^{\Phi} = {(p^2 + q^2)\over D}\eqn\nconfdila$$
The gauge fields are
$$A_\tau =  (p^2 + q^2 + P - Q ){\sinh \alpha \over D}\eqn\potcua$$
$$A_\sigma =  (P q^2 + p^2 Q ){\sinh \alpha \over D}\eqn\pottre$$
and the antisymmetric tensor is   
$$B_{\tau\sigma} = {b_0\over D} (P q^2 + p^2 Q )\eqn\antib$$

The axion field associated with  
antisymmetric field can be 
obtained from \lache\ and the equation for $H_{\mu\nu\rho}$, 
that gives rise to the expression 
$H^{\mu\nu\rho} = - {1\over \sqrt{-g}}e^{2 
\Phi} \epsilon^{\mu\nu\rho\sigma}\partial_\sigma \Psi$. 

After we have fixed the parameters $\epsilon = 1$ and $\lambda 
= 0$ in the structural functions $P$ and $Q$, one obtains
the scalar axion field
$$\Psi = 2 b_0 {(n q - m p)\over p^2 + q^2}\eqn\theaxion$$ 
From the solution presented above with $\gamma = 0 $ \chargesconf\ - \antib, one can derive the 
solution 
given in [\GGK], which has been obtained by direct integration of the equations of motion.

The asymptotic expansion of the metric, dilaton and electromagnetic 
potential allows to interprete new parameters corresponding to 
the mass, an scaled NUT-parameter, electric and magnetic charges, and 
dilaton and axion charges as  
$$ M = b_1 m; \qquad  e' = 2 m \sinh \alpha ; \qquad d_0 = b_0 m $$
$$ N = b_1 n; \qquad  g' = 2 n \sinh \alpha ; \qquad a_0 = b_0 n $$
where we have used the coordinate transformation 
$$ q \rightarrow  r ; \qquad p \rightarrow a \cos\theta - {2 b_0 m^2 
\over n}$$ 
leading to  expansions such as  
$$-g_{00} \sim 1 - {2 M\over r} - {2 N a \cos\theta\over r^2} + \ldots $$
$$A_0 \sim  {e'\over r} + {g' a \cos\theta\over r^2} + \ldots $$
$$e^\Phi \sim 1 + {2 d_0\over r} +  {2 d_0 a \cos\theta\over r^2} + \ldots $$
$$\Psi \sim {2a_0\over r} +\ldots $$

We can also rewrite \chargesconf\ as
$$ \eqalign{ds_E^2 &= {D\over P}
dp^2 + {P\over D}(d\tau + Q_\Phi d\sigma)^2 \cr
&+ {D\over Q}dq^2 - {Q\over D}(d\tau - P_\Psi
d\sigma)^2\cr}\eqn\nchargesconf$$
where
$$Q_\Phi = b_0 \gamma - 2 d_0 q + q^2; \qquad 
P_\Psi = - b_0 \gamma - 2 a_0 p + p^2$$
We have then obtained a general solution to the heterotic string equations of the action 
\action, 
that includes dilaton and axion fields. This solution contains all the known dynamical and 
kinematical parameters except the acceleration parameter. This parameter can presumably be 
included by 
adding a conformal factor ${1\over (1 - pq)^2}$ to \sinconformal\ then having \pdmetri\ as the  
starting metric to generate a more general solution.

Similarly as with their couterparts in the Einstein-Maxwell theory the electromagnetic field components
can be aligned with  the directions 
$$e_3 (e_4) = {1\over\sqrt{2}}\{({D\over Q})^{1/2} dq \pm ({Q\over D})^{1/2} (d\tau 
- P_\Psi d\sigma)\}$$  

In this case, as it has been presented in [\GGK] the values of the Weyl coefficients indicate that 
these kinds of gravitational fields are algebraically general as opposed to the type-D 
characteristic of the Einstein-Maxwell solution \sinconformal\ - \ncfweyl\ [\PLE]. 

\vfil\eject

\noindent {\bf 2.2 Dyonic solutions from type-D metrics without conformal 
factor}

In the solution previously presented the NUT parameter allows to generate 
a dyonic solution with 
electric and magnetic charge. One can also generate a new dyonic solution by  using the fact 
that the equations of motion of 
the action \action\ are invariant under an $SL(2, R)$ transformation [\SW, \sencua, \cliffo]. 
Then as for the case of electromagnetic field coupled to gravity their equations 
of motion are invariant under duality transformations in which the 
electric and magnetic fields are continously rotated into one another 
allowing solutions that carry both electric and magnetic charge.

The transformations \slf\ and \sllam\  give rise to a new electromagnetic field 
$$ \sqrt{1 + c^2} \hat f_{\mu\nu} = - (\Psi + c) f_{\mu\nu} + 
e^{-\Phi} \tilde f_{\mu\nu}\eqn\nuevof$$ 
and  new axion and dilaton fields [\sencua]
$$\hat \Psi = -{(1 + c^2)(\Psi + c) \over (\Phi + c)^2 + e^{-2\Phi}}\eqn\newaxi$$
$$e^{-\hat \Phi} = {(1 + c^2) e^{-\Phi}\over (\Phi + c)^2 + e^{-2\Phi}}\eqn\newdila$$ 
keeping the Einstein metric \nchargesconf\ invariant.
In the case we are dealing with, then 
$$\hat \Psi = - {p^2 - 2 d_0 p + q^2 + 2 a_0 q\over (p - d_0 - a_0)^2 + 
(q - d_0 + a_0)^2 } $$
$$e^{- \hat \Phi} = {p^2 - 2 a_0 p + q^2 - 2 d_0 q\over 
(p - d_0 - a_0)^2 + (q - d_0 + a_0)^2 } $$
where we have set  the parameter $c = 1$. They have very similar expressions 
with a new dilaton charge $a_0$ and an axion charge $-d_0$. 
From the electromagnetic fields \nuevof, whose expressions are more involved,  one 
can obtain a new electric charge 
$Q_E = 2(m + n)\sinh \alpha$  and a magnetic charge $Q_M = 2(m - n)\sinh \alpha$, 
provided that $m \neq n$.

\vfil\eject

\noindent {\bf 3.3 The C-metric and its stringy version}

We obtained the general solution that includes the solutions previously studied [\sentre, \KKOT, 
\cliffo, \GK, \GGK]. Now it would be interesting to include the conformal factor as in the case 
of the Einstein-Maxwell theory.
One of the simplest cases with this conformal factor is the so-called 
C-metric.
With  a contraction procedure the metric \pdmetri\ can be reduced to
a Kinnersley-Walker type of metric [\KW]
$$ \eqalign{ds^2 &= {1\over (p+q)^2} [{1\over P(p)} dp^2 
		    + {1 \over Q(q)}dq^2 + 
		 P(p) d\sigma^2 - Q(q) d\tau^2]\cr}\eqn\ccmetri$$
with 
$$P(p) = (\gamma - {\lambda\over 6}) + 2n p - \epsilon p^2 + 2m p^3 
- (e^2 + g^2)p^4\eqn\lape$$
$$Q(q) = - (\gamma + {\lambda\over 6}) + 2n q + \epsilon q^2 + 2m q^3 
+ (e^2 + g^2)q^4\eqn\laqu$$
with the electromagnetic field as follows
$$ f_+ = d ((e + ig )(q d\tau + i p d\sigma ))\eqn\cieinelec$$
and the invariant  
$${\cal F} = -\half(e + i g)^2(p + q)^4\eqn\cinvaele$$
The only nonvanishing component of the Weyl tensor is 
$$C^{(3)} = 2 m (p +q)^3 - 2 (e^2 + g^2)(p + q)^3(p - q)\eqn\ciweyl$$ 
Let us consider the case with $\lambda = n = e^2 + g^2 = 0$, which 
can be taken to the form  
$$ \eqalign{ds^2 &= {1\over A^2(x+y)^2} [{1\over F(y)} dy^2 
		    + {1 \over G(x)}dx^2 + 
		 G(x) dz^2 - F(y) dt^2]\cr}\eqn\cmetri$$
where $$G(x) = 1 - x^2 - 2m A x^3\eqn\lage$$
 $$F(y) = -1 + y^2 - 2m A y^3\eqn\laefe$$
This  space-time has been interpreted as the space-time of two particles 
with uniform acceleration $A$ [\KW].

The idea now is to construct a stringy version of this space-time by 
using the technique described above. After this $O(1, 1)$ transformation 
the charged solution obtained is given by 
$$ \eqalign{ds'^2 &= {1\over A^2(x+y)^2} [{1\over F(y)} dy^2 
		    + {1 \over G(x)}dx^2 +
      G(x) dz^2 - { A^4 (x + y)^4 \over D^2} F(y) dt^2]\cr}\eqn\cmetri$$
where 
$$  D = b_0 F(y) + b_1 A^2 (x+y)^2\eqn\lade$$
In this new configuration the dilaton field given by \dilato\ is 
$$e^{\Phi} = {A^2 (x + y)^2\over D} \eqn\cdilaton$$
Now, we pass on the expression of the new stringy metric as an Einstein 
metric $ds_{\rm E} = e^{-\Phi} ds'^2 $, thus
$$ \eqalign{ds_{\rm E}^2 &= {1\over A^2(x+y)^2} [{1\over e^\Phi F(y)} dy^2 
		    + e^{-\Phi}({1 \over G(x)}dx^2 +
      G(x) dz^2) - e^\Phi F(y) dt^2]\cr}\eqn\ecimetri$$
The nonvanishing electromagnetic potential is given by
$$A_t =  {A^2 (x+y)^2 - F(y)\over D} \sinh\alpha \eqn\potential$$
and the antisymmetric field remains zero.
The properties of this space-time will be presented elsewhere.


\vfil\eject

\noindent{\bf 3.  Summary}

In this work we have presented a general solution to the equations of 
motion of the heterotic string using duality transformations. 
Since the transformation that we have used [\sentre, \sencua] only modifies the 
dynamical parameters 
of the starting metric, namely the mass, the NUT-parameter, the electric  and 
magnetic charges, leaving the kinetical parameters unmodified, then it is possible to 
extend the known seven-parameter solution of the Einstein-Maxwell theory [\PD] to a new 
field configuration consistent with the presence of the dilaton and axion fields. In 
this case the new metric can also contain the parameter  of acceleration. 
We should notice that the electromagnetic field can be aligned with some 
null directions alike the case of the Einstein-Maxwell theory.

We have given a solution that contains the previous constructions [\sentre, 
\KKOT, \cliffo, \GK, \GGK] in which 
only some of the parameters were included. We have also given a particular solution 
that extends the so-called C-metric to the context of string theory.
 This kind of metric has been of 
interest in the study of pair-particle creation near an event horizon 
[\DGKT], as it can be 
interpreted as an space-time representing two particles with constant acceleration.    
This aspect will be explored further elsewhere.

\medskip

\noindent{\bf 4.  Acknowledgments}

I would like to thanks Alberto Garc{\'\i}a for useful discussions and Paule 
Dalens for her support  
and also wish to acknowledge the hospitality of the 
International Centre of 
Theoretical Physics, Trieste and CPT of \'Ecole Polytechnique
at some stages of this work.

\bigskip

\vfil\eject

\refout

\bye